\documentclass[preprint]{aastex}

\begin{document}

\title{Physical Properties of Main-Belt Comet P/2005 U1 (Read)
       \footnote{
       Some of the data presented herein were obtained at the W. M.
       Keck Observatory, which is operated as a scientific partnership
       among the California Institute of Technology, the University of
       California, and the National Aeronautics and Space Administration.
       The Observatory was made possible by the generous financial support
       of the W. M. Keck Foundation.
       Additionally, some data were obtained at the Gemini
       Observatory, which is operated by the Association of Universities for
       Research in Astronomy, Inc., under a cooperative
       agreement with the NSF on behalf of the Gemini partnership.}}

\author{Henry H. Hsieh$^{a,b}$, David Jewitt$^{a}$,
         and Masateru Ishiguro$^{c}$}

\affil{
    $^{a}$Institute for Astronomy, University of Hawaii, 2680 Woodlawn
    Drive, Honolulu, HI 96822, USA\newline
    $^{b}$Astrophysics Research Centre, Queen's University,
    Belfast, BT7 1NN, United Kingdom\newline
    $^{c}$Dept. of Physics and Astronomy, Seoul National University,
    Seoul, South Korea
}
\email{h.hsieh@qub.ac.uk, jewitt@ifa.hawaii.edu, ishiguro@astro.snu.ac.kr}

\slugcomment{Accepted, AJ, 2008-10-07}

\begin{abstract}
The main-belt comets
occupy dynamically asteroidal orbits in the main asteroid belt.
Here we present physical observations of the second-known member
of this population, P/2005 U1 (Read), which showed vigorous
cometary activity from 2005 October 24 to 2005 December 27.  Monte
Carlo numerical simulations of P/Read's dust emission indicate
that the coma and tail are optically dominated by dust particles
larger than 10~$\mu$m in size with terminal ejection velocities
of 0.2 to 3~m~s$^{-1}$.  We estimate P/Read's mass loss rate during
this period to be approximately 0.2~kg~s$^{-1}$, roughly an order
of magnitude larger than that calculated for 133P/Elst-Pizarro.
We also find that emission likely began at least
two months prior to P/Read's discovery, though we note this is a
lower limit and that earlier start times are possible.
Optical colors
measured for P/Read while it was active are approximately solar
($B-V=0.63\pm0.05$, $V-R=0.37\pm0.04$, $R-I=0.39\pm0.04$)
but are likely to be dominated by coma particles.  Observations of
P/Read in 2007 when it appears largely inactive show an extremely
small nucleus with an absolute magnitude of $H_R\sim20.1\pm0.4$,
corresponding to an effective radius of $r_{e}\sim0.3$~km.
P/Read's activity is consistent with sublimation-driven dust emission
and inconsistent with dust emission due to an impact, though 
the unusual strength of the 2005 outburst suggests the possibility that
it could have been due to the sublimation of a freshly-exposed
reservoir of volatile material.
\end{abstract}


\newpage

\section{INTRODUCTION}
Discovered on UT 2005 October 24 \citep{rea05}, P/2005 U1 (Read)
(hereafter P/Read) occupies an orbit in the main asteroid belt
($a=3.165$~AU, $e=0.253$, $i=1.267^{\circ}$).  It has a Tisserand
parameter (with respect to Jupiter) of $T_J=3.153$, while classical
comets have $T_J<3$ \citep{vag73,kre80}.  The orbit of this comet is
decoupled from Jupiter and is indistinguishable from
the orbits of ordinary main-belt asteroids, making it the second-known
member of the recently-identified population of main-belt comets
\citep[MBCs;][]{hsi06b}.

Three MBCs are currently known --- P/Read,
133P/Elst-Pizarro (also 7968 = 1999 N$_2$; hereafter 133P), and
176P/LINEAR (also 118401 = 1999 RE$_{70}$; hereafter 176P).
Discovered from limited observational data, these three objects imply a
true population of perhaps 150 currently active MBCs and likely many more
dormant MBCs \citep{hsi06b}.  Like P/Read, 133P and 176P also have $T_J>3$,
indicating that they are likewise dynamically decoupled from Jupiter.

While $T_J>3$ does not assure long-term dynamical
stability, it nonetheless strongly suggests that the three MBCs
are at least dynamically stable on their current orbits on
timescales much longer than those of other comets, and are not likely to be
recent arrivals from the outer solar system ({\it i.e.}, from the Kuiper
Belt or Oort Cloud).  This implication is further supported by numerical
simulations \citep[{\it e.g}.][]{ipa97,jfer02,lev06} which show that the
dynamical transition of comets from the outer solar system onto main belt
orbits should occur extremely infrequently, if at all, given the current
orbital configuration of the major planets.  The difficulty of executing
such a transition coupled with the likely abundance of MBCs instead suggests
that these objects are native members of the main asteroid belt, not
interlopers recently captured from elsewhere.  Finally,
recent thermal models show that subsurface ice on main-belt asteroids at
the heliocentric distance of the three MBCs
can in fact survive over billion-year timescales if protected from direct
sunlight by a dusty surface layer only a few meters in thickness
\citep{sch08}.
Such evidence suggests that ice could be widespread in
the main belt, though so far, actual cometary activity has only been
observed in the three currently-known MBCs.
Here, in an effort to shed further light on the nature
of these puzzling objects, we present new physical observations of P/Read.

\section{OBSERVATIONS}

Observations of P/Read were made in near-photometric conditions on multiple
nights in 2005 and 2007 (Table~\ref{obs_read}) using the University of
Hawaii (UH) 2.2~m telescope, the 8~m Gemini North telescope, and the
10~m Keck I telescope, all on Mauna Kea.  
Observations with the UH 2.2~m telescope were made using a Tektronix
2048$\times$2048 pixel CCD (image scale of $0\farcs219$ pixel$^{-1}$)
behind standard Kron-Cousins BVRI broadband filters.
Gemini observations were made using the Gemini Multi-Object Spectrograph
\citep[GMOS;][]{hoo04} (image scale of $0\farcs145$ pixel$^{-1}$
while using 2x2 binning) behind $g'r'i'z'$ filters \citep{fuk96}.
Observations with Keck were made using the red side of the
Low Resolution Imaging
Spectrometer \citep[LRIS;][]{oke95} in imaging mode.  The red side of
LRIS employs a Tektronix $2048 \times 2048$ CCD with an image scale of
$0\farcs210$~pixel$^{-1}$ and standard Kron-Cousins BVRI filters.

Standard image preparation (bias subtraction and flat-field reduction)
was performed.  For data from the UH 2.2~m and Gemini telescopes, flat
fields were constructed from dithered images of the twilight sky, while
images of the illuminated interior of the Keck I dome were used to construct
flat field images for Keck data.  Photometry of \citet{lan92} standard
stars and field stars was obtained by measuring net fluxes within circular
apertures, with background sampled from surrounding circular annuli.
Comet photometry was performed using circular apertures of different radii
(ranging from $2\farcs0$ to $5\farcs0$) but, to avoid the contaminating
effects of the coma, background sky statistics were measured manually in
regions of blank sky near, but not adjacent, to the object.  Several
(5 to 10) field stars in the comet images were also measured and used to
correct for extinction variation during each night.

\section{RESULTS\label{results}}

\begin{figure}
\plotone{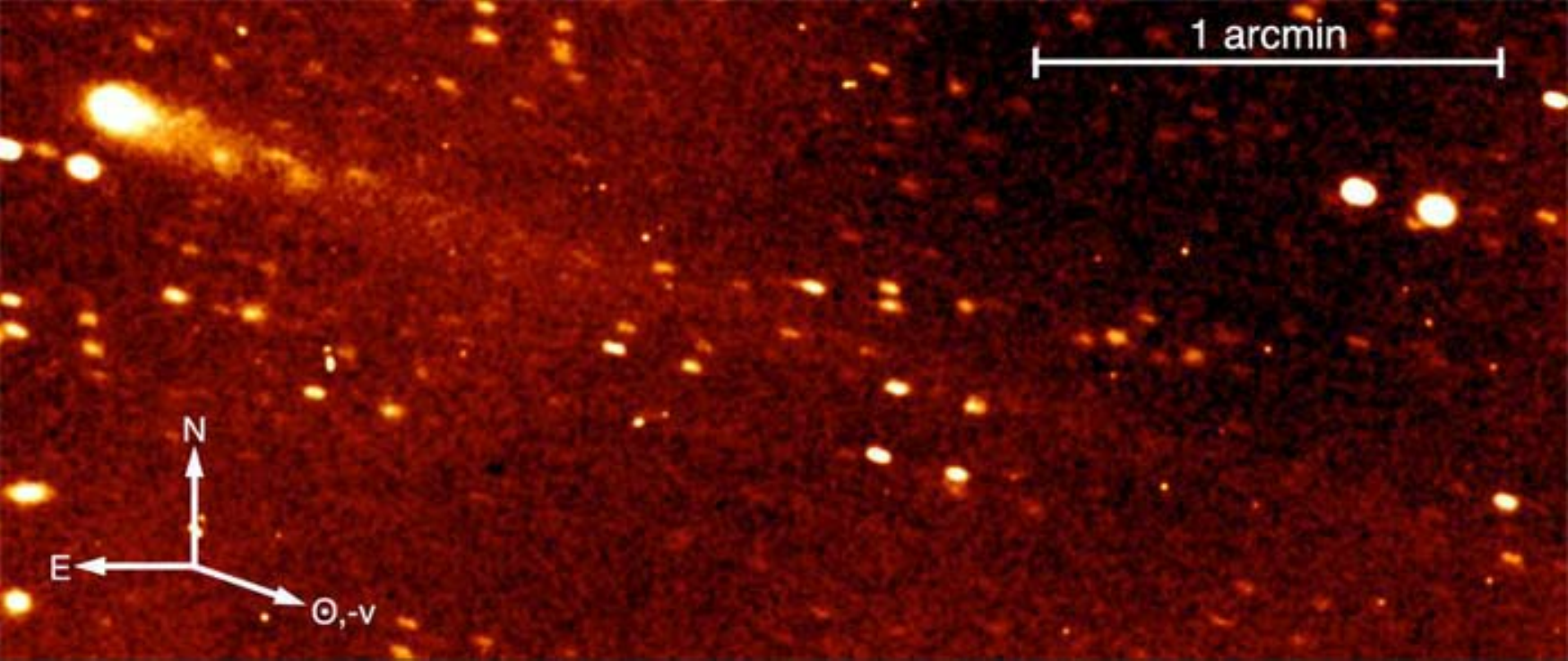}
\caption{\small Gaussian-smoothed composite r'-band image of P/Read
  (upper left corner) comprising 420~s of
  effective exposure time on the 8~m Gemini North telescope
  (equivalent to $\sim1.5$~hr on the UH 2.2~m telescope, scaling
  for different telescope aperture sizes), constructed from data
  obtained on UT 2005 November 26.  Gaussian smoothing has been applied
  to enhance the visibility of low surface brightness features,
  {\it i.e.}, P/Read's dust trail.
  Arrows indicate north (N), east (E),
  and the negative heliocentric velocity vector (--v) and the direction
  towards the Sun ($\odot$).
  Field stars are slightly elongated due to the the non-sidereal
  tracking of the object.}
\label{pread_gem}
\end{figure}

\begin{figure}
\includegraphics[width=5.0in]{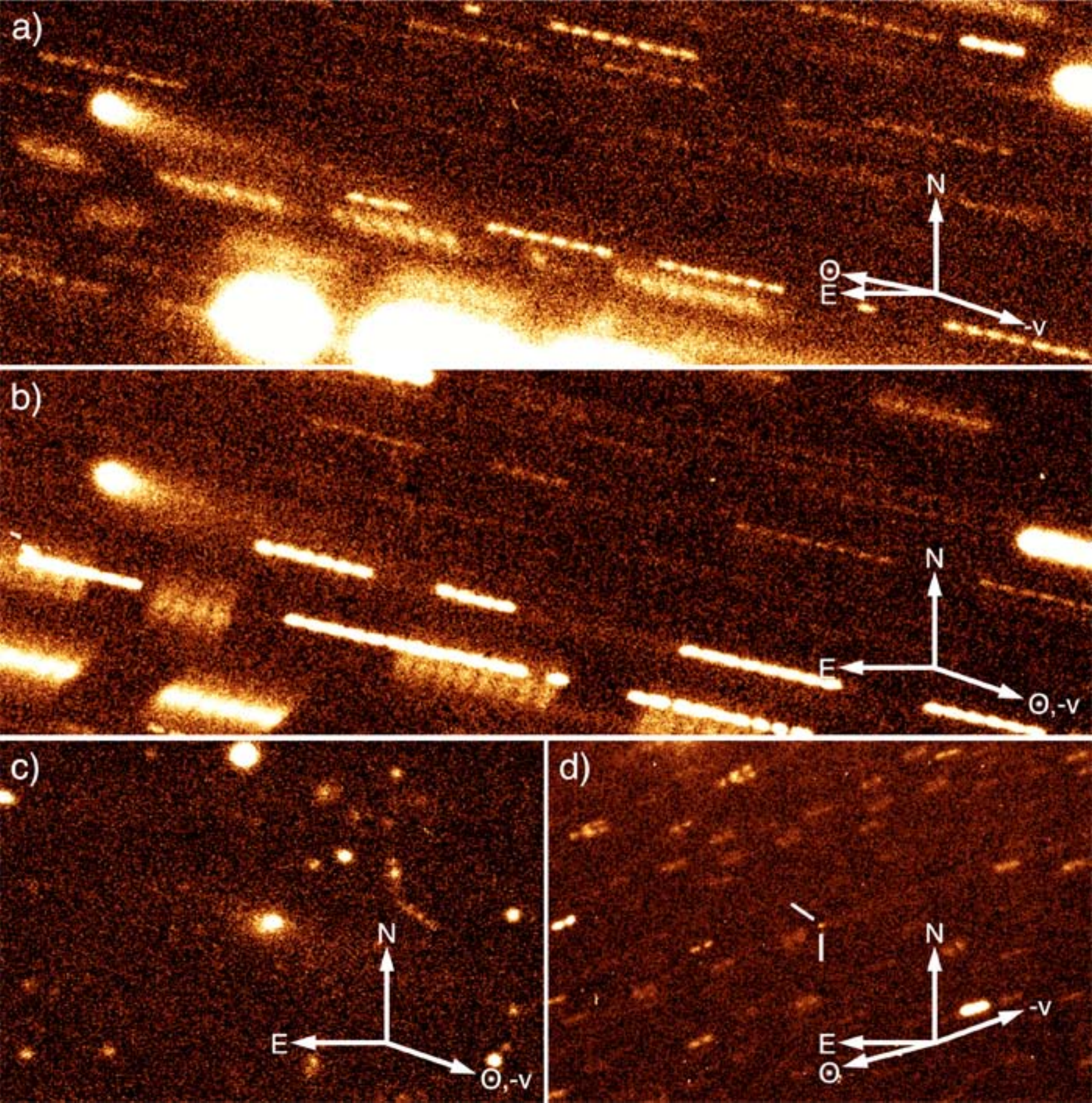}
\caption{\small Composite images of P/Read constructed using data from
  (a) UT 2005 November 10 (6900~s in $R$-band; UH 2.2~m telescope),
  (b) 2005 November 19 (15300~s in $R$-band; UH 2.2~m),
  (c) 2005 December 24 (1800~s in $R$-band; UH 2.2~m), and
  (d) 2007 January 27 (720~s in $R$-band; Keck I 10~m telescope).
  In (a) and (b), the comet nucleus is located in the upper left
  of each $1'\times3'$ panel.
  In (c) and (d), the comet nucleus is located in the center of each
  $1'\times1\farcm5$ panel.
  Arrows indicate north (N), east (E), and
  the negative heliocentric velocity vector (--v)
  and the direction towards the Sun ($\odot$).
  Dotted trails are the result of the non-sidereal tracking of the object
  and the summing of multiple individual exposures.
}
\label{pread_panels}
\end{figure}

P/Read was unambiguously active each time it was observed in
2005 (Figures~\ref{pread_gem},
\ref{pread_panels}a, \ref{pread_panels}b, and \ref{pread_panels}c).
No unambiguous activity was apparent when it was observed again in 2007
(Figure~\ref{pread_panels}d), though its faintness at the time makes a
rigorous assessment of its activity level difficult.
Composite images constructed from individual $r'$-band images
(aligned on P/Read's photocenter using linear interpolation)
obtained on UT 2005 November 26 using the Gemini North telescope 
(Figure~\ref{pread_gem}) and individual $R$-band images
obtained on UT 2005 November 10, 2005 November 19, and 2005 December 24
using the UH~2.2~m telescope (Figures~\ref{pread_panels}a,
\ref{pread_panels}b, and \ref{pread_panels}c, respectively)
all show P/Read possessing a strong coma
and dust tail closely aligned with the plane of its orbit as projected on
the sky.  This dust tail is seen in our data from November 26
to extend at least 3 arcmin before reaching
the edge of our detector's field of view, and points primarily to the
southwest throughout the month of November 2005.
This corresponds to the antisolar direction on 
November 10, but for November 19-22 and November 26, corresponds
to the projected direction towards the Sun (see Table~\ref{obs_read}).  By
December 24 and 25, portions of P/Read's tail
are seen to be projected in both the solar and antisolar directions.
These changes in the appearance of P/Read's dust tail during these
observations provide useful constraints on dust ejection
models (Section~\ref{dustmodeling}).

The coma that enveloped the nucleus of P/Read during the majority of
our observations makes it difficult to ascertain various physical
properties of interest.  For example, we searched for evidence of
rotational brightness modulation from time-series $R$-band photometry
over several nights using a phase dispersion minimization algorithm
\citep{ste78}.
We find candidate periods of 14.20~hr and 17.82~hr, and secondary
candidate periods of 7.32~hr, 8.06~hr, and 10.29~hr.  None of these
solutions, however, produces a coherent lightcurve when used to phase
the photometry. We attribute this to the dilution of the light from the
nucleus by light scattered from near-nucleus dust, to the sensitivity
of measured coma magnitudes to the seeing, and to the variation of that
seeing from night to night and often even between consecutive exposures.
We therefore regard our attempt to discern P/Read's rotational period as
unsuccessful. Detailed follow-up observations and analysis at a future
time when P/Read is observed to be inactive would be useful.

From multi-filter photometry on several nights, we find mean colors of
$B-V=0.63\pm0.05$, $V-R=0.37\pm0.04$, and $R-I=0.39\pm0.04$.  These are
comparable to solar colors as well as to the colors of other active comets,
inactive comet nuclei, and the nearly bare nucleus of the first known MBC,
133P (Table~\ref{colors}).
P/Read's colors are also consistent with those of
C-type asteroids which dominate the Themis family \citep{flo99}
in which the other two MBCs (133P and 176P) are found.
Color measurements remain largely constant with increasing
aperture radii (and therefore increasing coma contribution)
up to $5\farcs0$, however, indicating that measurements at
all radii are likely dominated by coma particles.

\begin{figure}
\includegraphics[width=6.0in]{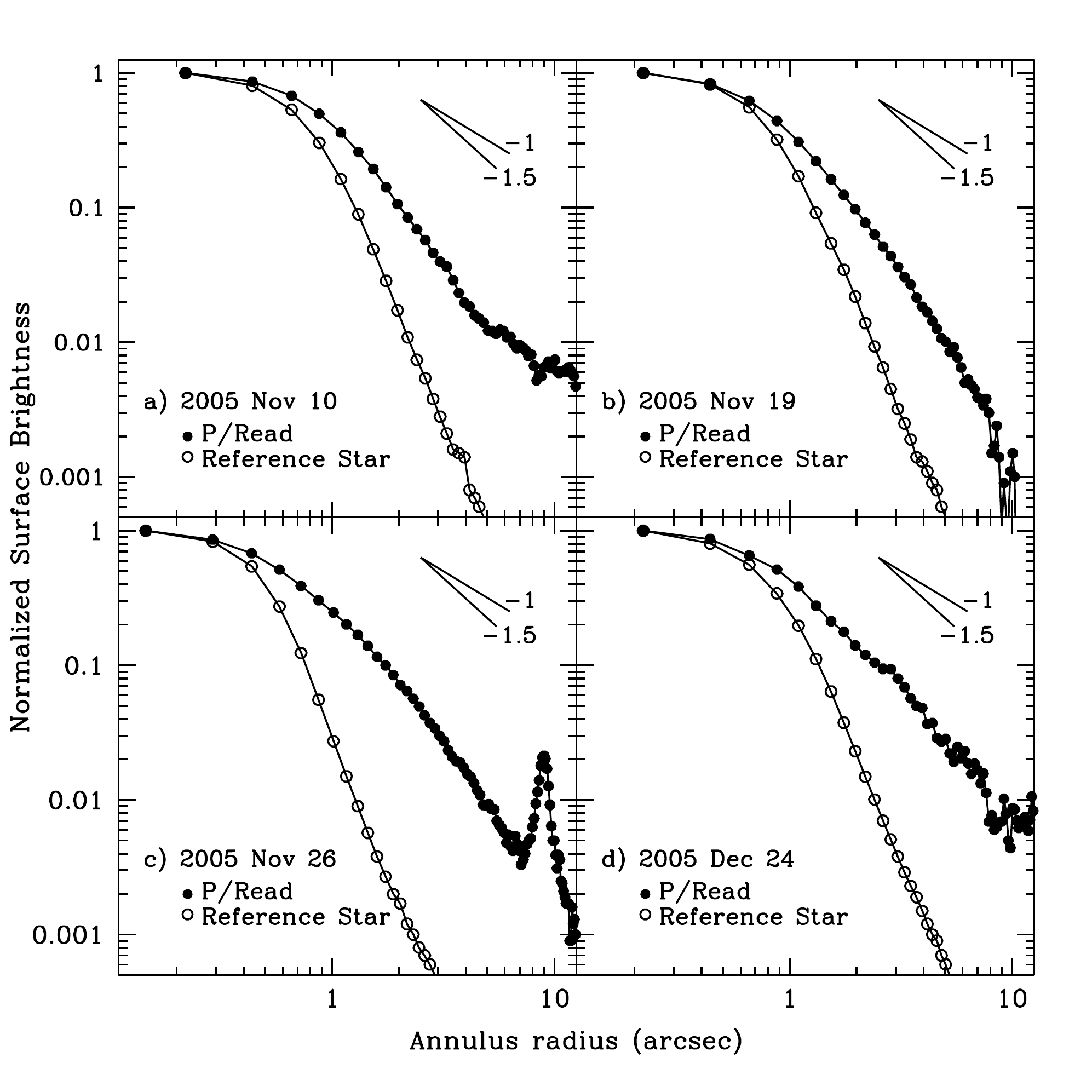}
\caption{\small Radial surface brightness profiles of P/Read
  and a reference star, from $R$-band data obtained on
  (a) 2005 November 10 ({\it cf.} Figure~\ref{pread_panels}a),
  (b) 2005 November 19 ({\it cf.} Figure~\ref{pread_panels}b), and
  (d) 2005 December 24 ({\it cf.} Figure~\ref{pread_panels}c) on the UH 2.2~m telescope,
  and
  (c) $r'$-band data obtained on 2005 November 26 ({\it cf.} Figure~\ref{pread_gem})
  on the Gemini North telescope.
  The surface brightnesses of the comet in the central aperture (radius of 1 pixel) are
  (a) $\Sigma=21.0$~mag~arcsec$^{-2}$,
  (b) $\Sigma=21.0$~mag~arcsec$^{-2}$,
  (c) $\Sigma=21.1$~mag~arcsec$^{-2}$, and
  (d) $\Sigma=22.2$~mag~arcsec$^{-2}$,
  while at 4~arcsec from the photocenter, surface brightnesses fall to
  (a) $\Sigma=25.4$~mag~arcsec$^{-2}$,
  (b) $\Sigma=25.4$~mag~arcsec$^{-2}$,
  (c) $\Sigma=25.6$~mag~arcsec$^{-2}$, and
  (d) $\Sigma=25.7$~mag~arcsec$^{-2}$.
  For all field star profiles and within $\sim7''$ from the nucleus for comet profiles,
  uncertainties are comparable to the size of the points plotted.
  Beyond $\sim7''$ from the nucleus for comet profiles, uncertainties
  can be visually estimated from the
  scatter of points from a smoothly-varying profile.
  Straight lines with slopes of $-1$ and $-1.5$, as marked, have
  been included for reference.
  }
\label{radprof_read}
\end{figure}

To examine the non-stellar nature of P/Read, we construct normalized radial
surface brightness profiles for P/Read (Figure~\ref{radprof_read})
from the composite images shown in Figures~\ref{pread_gem} and
\ref{pread_panels} (excluding the image shown in Figure~\ref{pread_panels}d
in which the object is too faint for this level of analysis)
and compare them to similarly-constructed profiles of field stars from
sidereally-tracked images taken on the same night at similar airmasses.  
The comparison of non-simultaneous data is not ideal but unavoidable as all
observations of P/Read were conducted using non-sidereal tracking on each
telescope to follow the comet, causing field stars to appear trailed and
therefore unsuitable for radial profile analysis.  As can be seen in
Figure~\ref{radprof_read}, while nightly seeing differences cause
fluctuations in the profiles of our comparison stars, the overall shape
of P/Read's coma profile remains essentially constant in November,
exhibiting minimal sensitivity to nightly seeing variations and changes
in viewing geometry ({\it e.g.}, solar phase angle), while flattening
slightly in December as the activity apparently weakens.

In order to estimate the size of P/Read's nucleus,
we scale the peak brightnesses of several sidereally-tracked field
stars to that of a stacked image of P/Read from UT 2005 Nov 10, measure
the flux contained within apertures centered on the photocenters of the
field stars and P/Read, and assume that the excess present in the broader
profile of P/Read is due to coma.  Using this procedure, we estimate that
the comet's nucleus contributes approximately 40\% of the total brightness
measured within an aperture $4\farcs0$ in radius.  This gives an approximate
nucleus magnitude of $m(R,\Delta,\alpha)\sim20.3$, or an
absolute magnitude
of $H_R\approx m(1,1,0)\sim17.5$,
where the solar phase angle, $\alpha$, is close to
zero at the time of the observations, obviating the need to assume a
phase-darkening function.
Then, using
\begin{equation}
p_Rr_e^2 = 2.24\times10^{22} 10^{0.4[m_{\odot}-m(1,1,0)]}
\label{effradius}
\end{equation}
\citep{rus16}, where $p_R$ is the geometric $R$-band albedo
and $m_{\odot}=-27.07$~mag
\citep{har80,har82,har90} is the apparent solar $R$-band magnitude,
we estimate an effective nucleus radius of $r_e\sim0.9(0.04/p_R)^{1/2}$~km.
Using an identical analysis for the dust component, we obtain a cumulative
absolute magnitude of $m(1,1,0)\sim17.1$ for the dust contained within a
$4\farcs0$ aperture (4200~km in the plane of the sky at the distance of
P/Read), corresponding to a total geometric scattering cross-section of
$C\sim\pi r_e^2\sim3.8$~km$^2$, assuming an optically thin coma.
Ignoring opposition surge effects (on which we currently
possess no useful constraints),
we derive an approximate total mass of visible dust within our $4\farcs0$
aperture of $m_d\sim\rho\bar{a}C\sim3.8\times10^4$~kg, assuming a
bulk grain density of $\rho=1000$~kg~m$^{-3}$ and an average grain size
of $\bar{a}=10~\mu$m.

This crude method of estimating the contribution of the nucleus to the total
measured flux of a comet possessing a strong coma has its deficiencies,
of course.  The computed contribution of P/Read's nucleus to the comet's
total measured brightness derived using this method is affected by
several complicating factors such as variations in seeing conditions
between the times when comet images are obtained to the times when
template stars are observed, and the unknown (but certainly non-zero)
contribution of
the coma to the central pixel of the comet (meaning that by scaling
field stars to this central brightness, we are already
overestimating the true nucleus contribution).
Nonetheless, this analysis produces relatively consistent results
when applied to data obtained on other nights in 2005, giving absolute
magnitudes for P/Read's nucleus of $H_R\sim17.4$ on UT 2005 Nov 19,
$H_R\sim17.3$ on UT 2005 Dec 24, and $H_R\sim17.3$ on UT 2005 Dec 25
(assuming a linear phase function with a phase-darkening coefficient of 0.035),
corresponding to an effective nucleus size of $r_e\sim1.0$~km
(assuming an albedo of $p_R=0.04$).
As with P/Read's rotational period, 
however, a definitive nucleus size can only really be expected from
observations of P/Read while it is inactive.

We obtained just such observations with the 10~m Keck I telescope in
2007 Jan 27 when P/Read was almost 1~AU farther from the
Sun, and presumably far less active.  No evidence of a coma is seen in
these observations, though the faintness of the nucleus makes
the existence of a 
coma difficult to definitively rule out.  We measure an apparent $R$-band
magnitude for the nucleus of $m(R,\Delta,\alpha)=24.9\pm0.4$
(Table~\ref{obs_read}).  Assuming a
phase-darkening coefficient of 0.035, this corresponds to an absolute
magnitude of $m(1,1,0)\sim20.1$ and an effective nucleus radius of
0.3 km, significantly smaller than our previous nucleus size
estimate.  Working backwards, we find that a nucleus of this size
would have constituted only 4\% of the total flux measured for an active
P/Read on 2005 Nov 10 instead of 40\% as determined from our analysis
of the coma's surface brightness profile at the time.
The size of the disparity between
these nucleus size estimates hints that it may be due to more than just
measurement uncertainties.  We discuss possible explanations for this
discrepancy in Section~\ref{discussion}.

\section{DUST CLOUD MODELING\label{dustmodeling}}

In order to obtain a more quantitative picture of P/Read's activity,
we numerically model its dust emission.  We acknowledge from the
outset that our model is unavoidably underconstrained and that any
results will be far from exact analytical descriptions of the comet's
dust emission.  However, our objective is to simply place constraints on
certain key properties such as grain size distribution, ejection velocities,
and the temporal nature of the emission.  As such, our modeling strategy is
formulated to focus on these key properties, while omitting
unconstrained second-order parameters
({\it e.g.}, the number, location, and directionality of jets, or the rate
and orientation of nucleus rotation)
and incorporating other assumptions in the interest of simplifying both
computation and interpretation of results.

The trajectory of a dust particle of radius, $a$, largely depends on the
ratio, $\beta$, of the particle's acceleration due to solar radiation
pressure to its acceleration due to
gravity, and on ejection velocity, $v_{ej}$ \citep{fin68}.
Syndyne curves are lines representing constant
values of $\beta$ when the ejection velocity is assumed to be zero, whereas
synchrone curves are lines representing particles ejected at the same time
and having a range of $\beta$ values.  Together, syndyne and synchrone
curves are often used to determine the range of $\beta$ values, which is
related to the range of particle sizes, of ejected dust from comets.
Since P/Read's low inclination means that it is always found close to the
ecliptic plane, where syndynes tend to overlap, it is difficult
to discern particle sizes from syndyne-synchrone analyses.  In addition,
even in high-inclination cases, syndyne-synchrone analyses can result in
misleading $\beta$ values \citep{ful04}.
Analysis allowing for non-zero ejection velocities is
essential for estimating the particle sizes of the dust emitted by P/Read
\citep[{\it cf}.][]{ish07}.

We assume that dust particles are ejected in cone-shaped jets that are
radially symmetric with respect to the Sun-comet axis with a half-opening
angle, $w$. The terminal velocity, $v_{ej}$, of the ejected dust particles
can be estimated from
\begin{equation}
v_{ej}(r_h,\beta) = v_0 \beta^{u_1}\left({r_h\over{\rm AU}}\right)^{-u_2}
\label{termvel}
\end{equation}
where $\beta$ is the ratio of solar radiation pressure to the gravitational
force on a particle, $r_h$ is the heliocentric distance,
$v_0$ is the reference
ejection velocity in m~s$^{-1}$ of particles with $\beta=1$ at $r_h=1$~AU,
and $u_1$ and $u_2$ are the power indices of the reference ejection velocity
dependence on $\beta$ and $r_h$.
We use an exponential size distribution with an index of $q$, and assume that
the dependence of dust production rate on heliocentric distance can expressed
by a simple exponential function with an index of $k$.  

\begin{figure}
\plotone{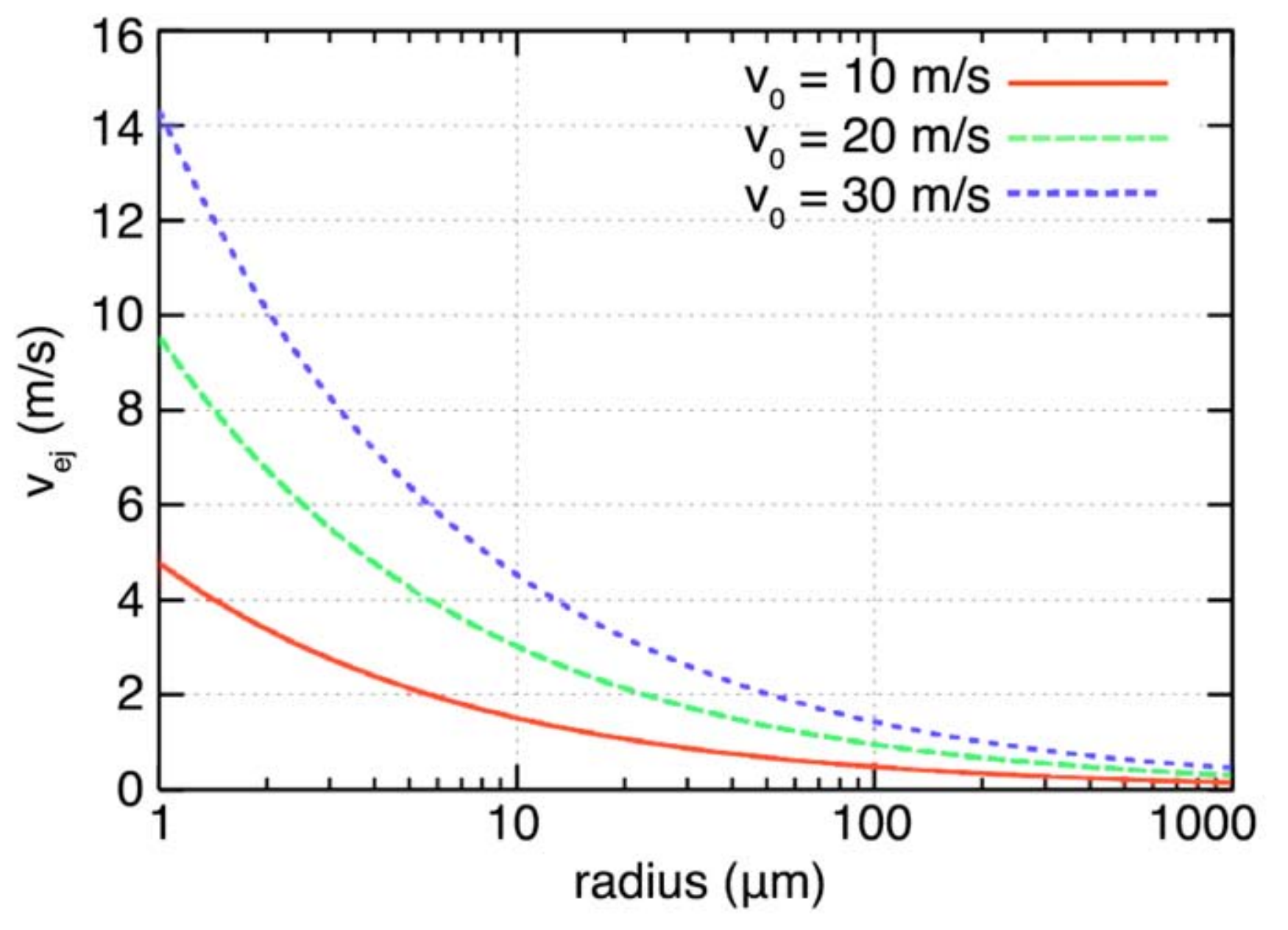}
\caption{\small Terminal ejection velocity ($v_{ej}$) as a function of
  particle size ($a$) at a heliocentric distance of $r_h=2.5$~AU, as given
  by Equation~\ref{termvel} using power indices $u_1=0.5$ and $u_2=0.3$.
  }
\label{ejectionvel}
\end{figure}

Dust ejection is assumed to begin two aphelion passages prior to the current
perihelion passage ({\it i.e.}, 1.5 orbits ago).  This is an assumption made
to simplify our initial analysis and is not intended to represent physical
reality.  In practice, the comet's coma and tail is dominated by
recently-ejected dust, and so emission much earlier in the
past has little effect on our final results.
Images of model comets
are generated by Monte Carlo simulations parameterized by $\beta_{min}$,
$\beta_{max}$, $v_0$, and $w$, and using fixed, typical values for
$u_1$, $u_2$, $q$, and $k$ 
\citep[$u_1=0.5$, $u_2=0.3$, $q=3.5$, and $k=3.0$;
{\it cf.}][]{lis98,rea00,ish07,sar07}.
Using these parameters,
and Equation~\ref{termvel}, 
we can compute terminal ejection velocities as functions of particle size at
$r_h=2.5$~AU where P/Read is located at the time of our observations
(Figure~\ref{ejectionvel}), and then calculate
apparent dust particle positions for a given
observing geometry. Pixel intensities, $I_{pixel}$,
at given CCD coordinates ($x$,$y$) are then given by
\begin{equation}
  I_{pix}(x,y) = \int_{{t_0}}^{t_{obs}}\int_{a_{min}}^{a_{max}}
                   F_{\odot}\left({r_h\over{\rm AU}}\right)^{-2}
		   \sigma(a,\alpha) N_{cal}(a,t)~da~dt
\label{intensity}
\end{equation}
where $t_0$ is the start time of the model simulation, $t_{obs}$ is the time
of observation, $a_{min}$ and $a_{max}$ are the minimum and maximum particle
sizes, respectively,
$F_{\odot}$ is the $R$-band
($\lambda=0.64$~$\mu$m) solar flux density
($1.60\times10^3$~W~m$^{-2}{\rm{\mu}m}^{-1}$)
at 1~AU, $r_h$ is the heliocentric distance, $N_{cal}(a,t)~da$ is the number
of dust particles with size $a$ to $a+da$, 
and the differential scattering cross-section, $\sigma(a,\alpha)$,
is given by
\begin{equation}
\sigma(a,\alpha) = {G\over\pi}A_p(\alpha)
\label{crosssec}
\end{equation}
where $G=\pi a^2$ is the geometric cross-sectional area of the particle, and
$A_p(\alpha)$ is the modified geometric albedo at the
phase angle, $\alpha$ \citep{han81}.

\begin{figure}
\includegraphics[width=5.0in]{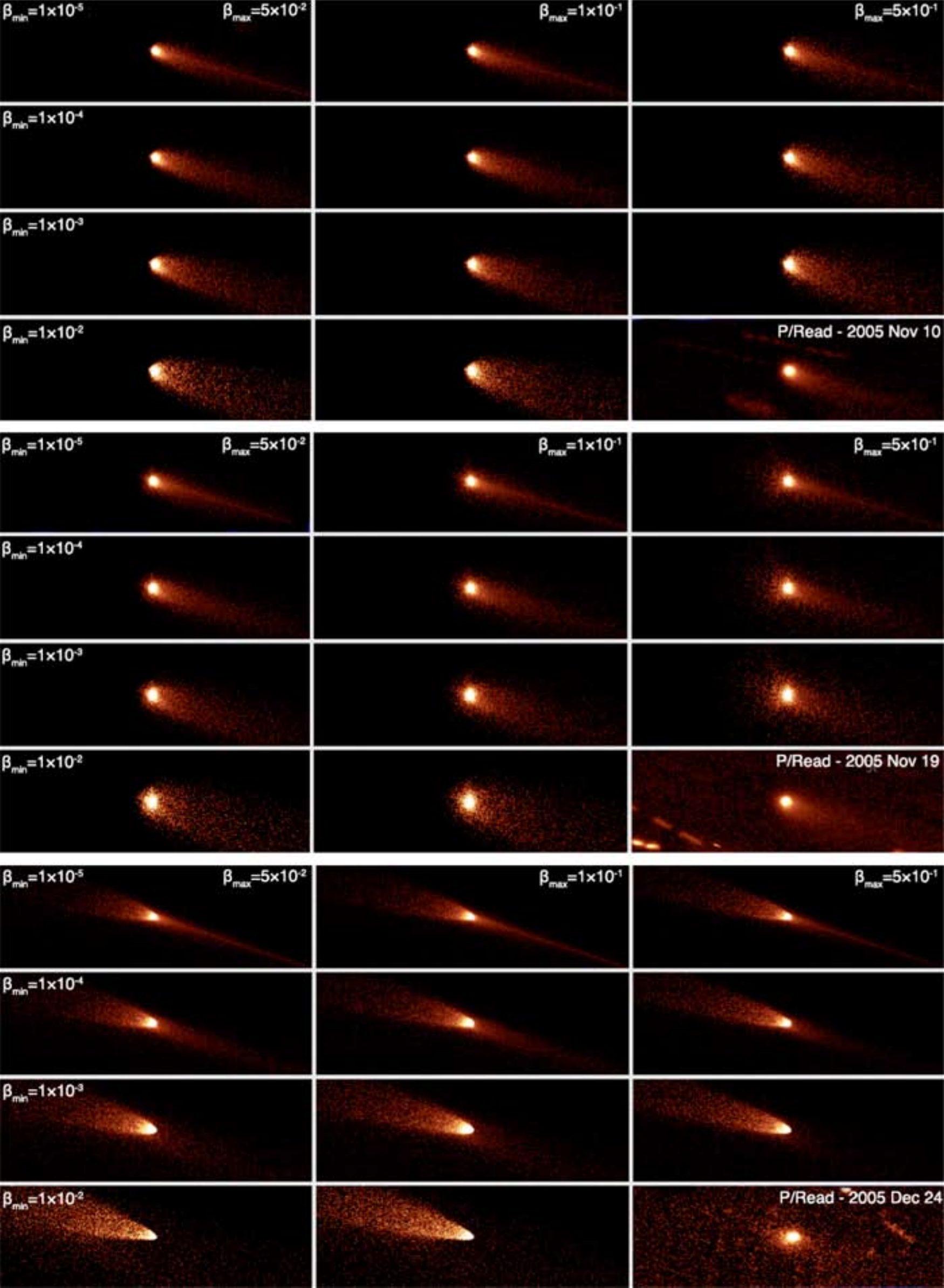}
\caption{\small Images of modeled dust emission for P/Read for
  UT 2005 November 10 (top group of panels), 2005 November 19
  (center group of panels), and 2005 December 24 (bottom group of panels).
  In these models, jet opening angles and reference ejection velocities are
  held constant at $w=45^{\circ}$ and $v_0=25$~m~s$^{-1}$, respectively.
  Minimum $\beta$ values and maximum $\beta$ values are varied between
  $1\times10^{-5} < \beta_{min} < 1\times10^{-2}$ and
  $5\times10^{-2} < \beta_{max} < 5\times10^{-1}$, respectively.
  Identical minimum $\beta$ values are used for models arranged in the same
  horizontal row while
  identical maximum $\beta$ values are used for models arranged in the same
  vertical column in each group of panels.
  }
\label{fpmodel_case1}
\end{figure}

\begin{figure}
\includegraphics[width=5.0in]{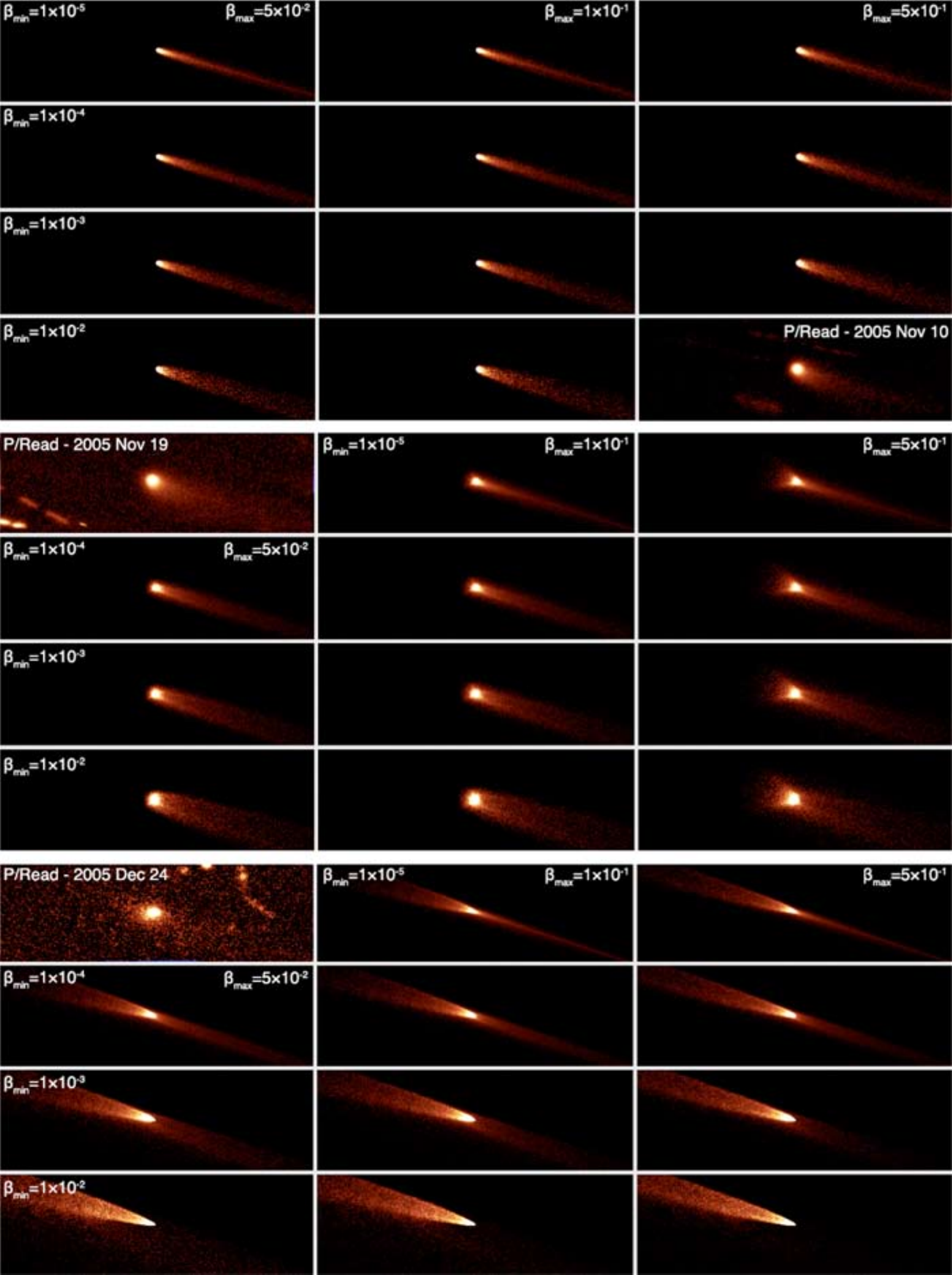}
\caption{\small Images of modeled dust emission for P/Read for
  UT 2005 November 10 (top group of panels), 2005 November 19
  (center group of panels), and 2005 December 24 (bottom group of panels).
  In these models, jet opening angles and reference ejection velocities are
  held constant at $w=90^{\circ}$ and $v_0=10$~m~s$^{-1}$,
  respectively.  Minimum $\beta$ values and maximum $\beta$ values are
  varied between $1\times10^{-5} < \beta_{min} < 1\times10^{-2}$ and
  $5\times10^{-2} < \beta_{max} < 5\times10^{-1}$, respectively.
  Identical minimum $\beta$ values are used for models arranged in the same
  horizontal row while identical maximum $\beta$ values are used for models
  arranged in the same vertical column in each group of panels.
  }
\label{fpmodel_case2}
\end{figure}

Multiple simulations are carried out using various parameter sets, and the
resulting model images are then visually compared to the data to find
plausible model parameters.  Chi-squared fitting of contour maps of the best
visual matches to contour maps of observed data is then used to find the
most plausible set of parameters from among our choices of reasonable visual
matches.
A full list of parameters tested is shown in Table~\ref{fpmodel_params}.
We identify two sets of plausible model parameters.
In the first (Figure~\ref{fpmodel_case1}), we find that for
$v_0=25$~m~s$^{-1}$, large particles
($\beta_{min}\gtrsim10^{-4}$, $\beta_{max}\lesssim10^{-1}$)
give rise to model images that closely match observations.  In this case,
the given $\beta$ values correspond to terminal ejection velocities of
$v_{ej}=0.2-1.9$~m~s$^{-1}$.
In the second set of models
(Figure~\ref{fpmodel_case2}), we find that calculations using smaller
particles ($\beta_{min}\gtrsim10^{-3}$,  $\beta_{max}\lesssim10^{-1}$)
ejected with a somewhat slower $v_0$
($v_0=10$~m~s$^{-1}$, corresponding to $v_{ej}=0.2-2.4$~m~s$^{-1}$)
also produce reasonable fits to the data.  Due to the low inclination of
P/Read, we are unable to distinguish between these two cases, but we
nonetheless conclude that particle sizes are certainly larger than 10~$\mu$m
and probably larger than 100~$\mu$m, and terminal ejection velocities are
approximately 0.2--3~m~s$^{-1}$.  
For this range of particle sizes and ejection velocities,
we compute an approximate mass loss rate of $dm/dt\sim0.2$~kg~s$^{-1}$.
This mass loss rate is
roughly an order of magnitude larger than that calculated for 133P
\citep[$dm/dt\sim2\times10^{-2}$~kg~s$^{-1}$;][]{hsi04}, consistent with
P/Read's much more active appearance.

The modeling described above assumes continuous dust emission
({\it i.e.}, in a manner consistent with dust emission driven by the cometary
sublimation of volatile material), an assumption based on P/Read's classical
cometary appearance.  The fact that this modeling is able to successfully
reproduce the observed morphology of P/Read strongly suggests that
the observed activity is accurately characterized by continuous
dust emission.  We can, however, address this presumption directly by
examining the longevity of P/Read's dust emission.

We compute phase-angle-corrected $Af\rho$ values
\citep[][Table~\ref{afrho}]{ahe84}
for each of our observations from 2005 using
\begin{equation}
Af\rho = {(2R\Delta)^2\over \rho} 10^{0.4(m_{\odot}-m_{obs})}
\end{equation}
where $R$ is in AU, $\Delta$ is in cm, $\rho$ is the physical radius in cm
of the photometry aperture at the distance of the comet, $m_{\odot}=-27.07$
is the solar $R$-band magnitude, and $m_{obs}$ is the phase-angle-corrected
$R$-band magnitude of the comet inside a $4\farcs0$-radius aperture.  We
find some fluctuation in measured values but note that the level of activity
remains reasonably steady for the entire 1.5 months between 2005 November 10
and 2005 December 25, with an average $Af\rho$ of 8.0~cm.

\begin{figure}
\plotone{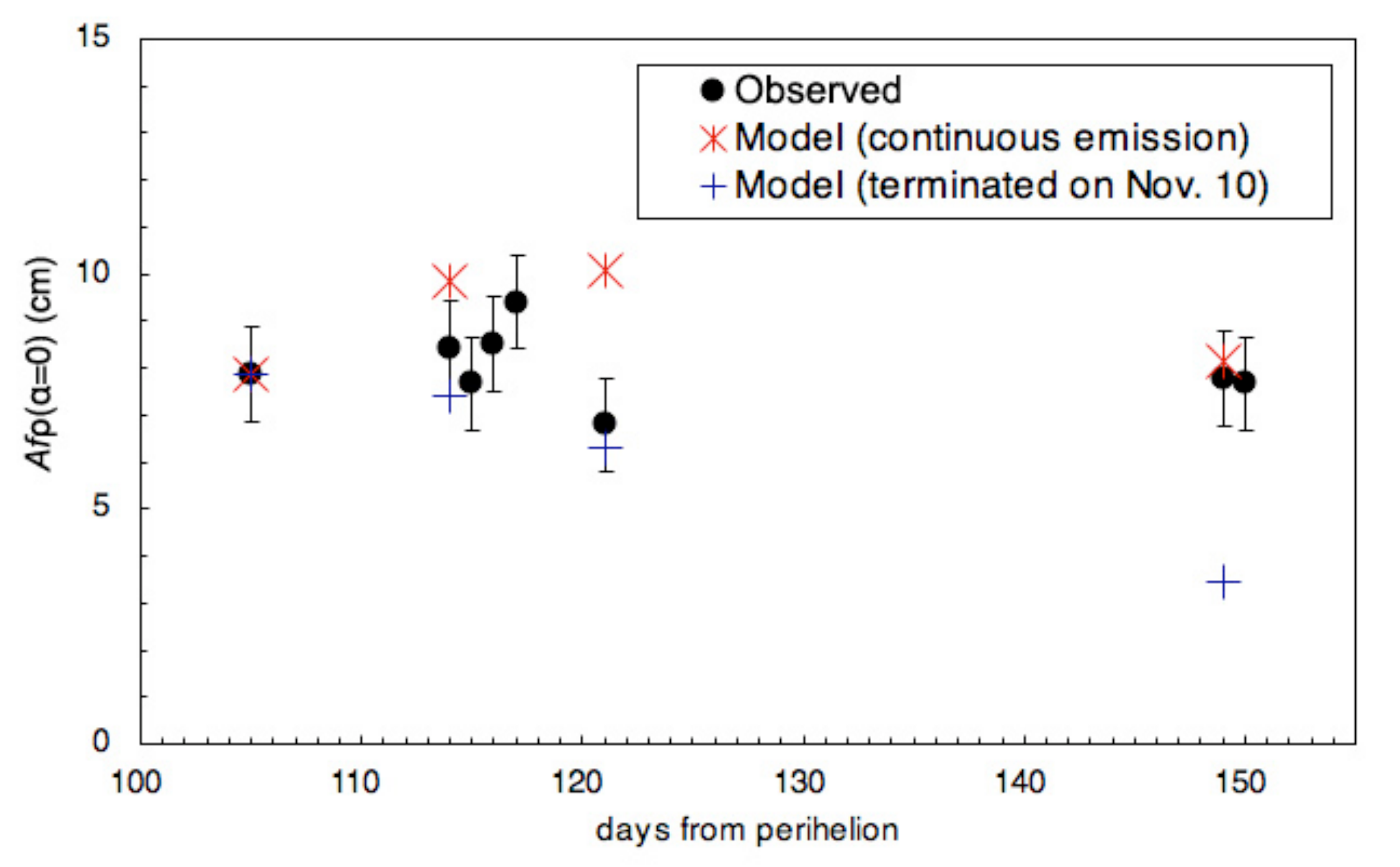}
\caption{\small Comparison of activity intensity, $Af\rho(\alpha=0)$, of observations
  and models of continuous emission and emission terminated on November 10.
}
\label{afrho_model}
\end{figure}

To investigate what this roughly constant level of activity implies for
P/Read's emission behavior, we perform a test on our best-fit model where we
terminate the dust emission as of 2005 November 10 (the date of our first
observation of P/Read).  We then repeat the analysis performed for our observed
data by computing phase-angle-corrected $Af\rho$ values for both this truncated
emission model and our original continuous emission model on selected dates,
normalizing both models to have the same $Af\rho$ values as the observed data
on 2005 November 10.
Results of this analysis are tabulated in Table~\ref{afrho} and
plotted in Figure~\ref{afrho_model}.  As can be seen from the tabulated and
plotted data, the activity level of the truncated emission model falls
well below the observed data by 2005 December 24.  The inconsistency of a
truncated emission model with observations indicates that P/Read was likely
actively emitting dust throughout our 2005 observations.  Such sustained dust
emission behavior over a long time period strongly suggests that P/Read's
activity was driven by the sublimation of volatile ices, {\it i.e.},
that it was cometary in nature.

\begin{figure}
\includegraphics[width=5.1in]{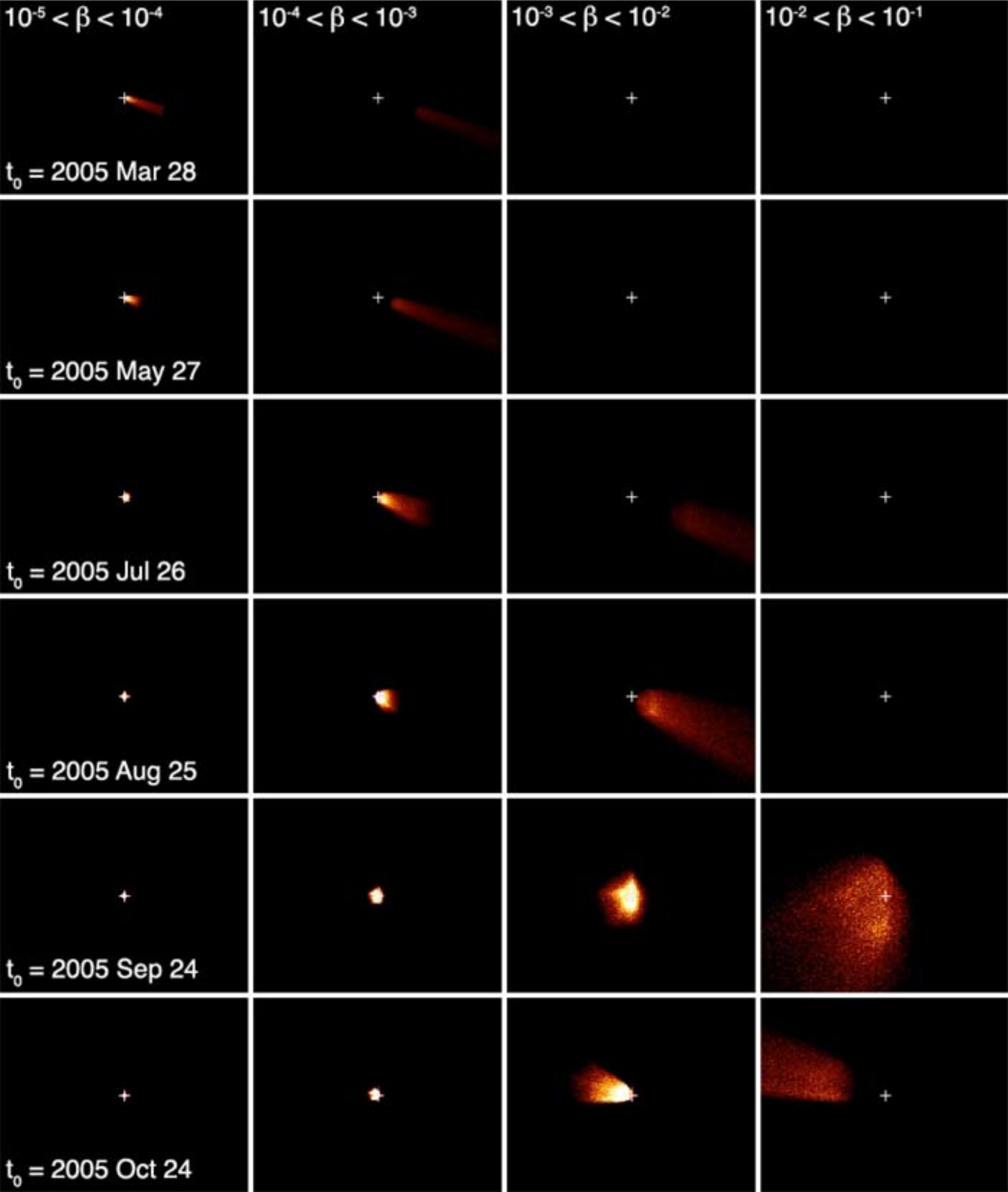}
\caption{\small Images of modeled impulsive dust emission events for P/Read
observed on UT 2005 December 25. In these models, jet opening angles and
reference ejection velocities are held constant at $w=45^{\circ}$ and
$v_0=25$~m~s$^{-1}$, respectively.
Dust released in single impulsive emission events on
2005 March 28 (272 days prior to observations on 2005 December 25),
2005 May 27 (212 days prior to observations),
2005 July 26 (152 days prior to observations),
2005 August 25 (122 days prior to observations),
2005 September 24 (92 days prior to observations), and
2005 October 24 (62 days prior to observations) are modeled,
and different ranges of $\beta$ values
are tested, from $10^{-5}<\beta<10^{-4}$ to $10^{-2}<\beta<10^{-1}$.
Identical ejection dates are used for models arranged in the same horizontal
row, while identical $\beta$ value ranges are used for models arranged in the
same vertical column.  Each panel is approximately 45~arcsec by 60~arcsec.
For $10^{-5}<\beta<10^{-4}$ panels in which no extended dust emission is
visible, dust is still concentrated near the nucleus (marked with a white
cross in each panel).  For all other panels in
which no activity is visible, dust has diffused beyond the field of view.
}
\label{impulse_model}
\end{figure}

To further validate this conclusion, we investigate whether impulsive
dust emission events (such as the formation of an ejecta cloud resulting from
the impact of another asteroid on P/Read's surface) could explain P/Read's
activity.  We do so by modeling the release of dust at a single time, $t_0$,
and then simulating the comet's expected appearance on 2005 December 25
(Figure~\ref{impulse_model}).  For this series of models, we hold all
parameters fixed except for $t_0$, $\beta_{min}$, and $\beta_{max}$, varying
$t_0$ from 2005 March 28 (272 days prior to the date of observations) to
2005 October 24 (the date of the discovery of P/Read's activity, and
62 days prior to our December observations), and varying $\beta$ value
ranges from $10^{-5}<\beta<10^{-4}$ to $10^{-2}<\beta<10^{-1}$.
We find that, regardless of particle size, dust emitted in a single
impulsive event tends to appear on only one side of the nucleus
when observed on 2005 December 25, with dust emitted on 2005 August 25 or
earlier forming a fan-shaped tail directed to the southwest and dust
emitted on 2005 October 24 forming a tail directed to the northeast.
An exception is emission on 2005 September 24 which results in a dust
tail aligned very closely to the line of sight on 2005 December 25, thus
appearing as a nearly-circular cloud superimposed on the nucleus and slightly
offset to the southeast.

Our observations on 2005 December 25 show dust features extending both to
the northeast and southwest of the nucleus (Figure~\ref{pread_panels}c).
Based on the results of our impulsive emission tests, we therefore conclude
that the northeast extension of P/Read's dust cloud must be due to particles
ejected after September 2005, while the southwest dust feature must be due
to particles ejected prior to September 2005.  Interestingly, since our
models show that particles with $\beta\sim0.1$ emitted at this time will have
dispersed well beyond the nucleus by 2005 December 25, the observed
southwest extension must actually be composed only of larger particles
($\beta\lesssim10^{-2}$).  Our main finding, however, is that no single
impulsive emission event can simultaneously produce dust features extending
in both directions.  At least one emission episode from both time periods
is necessary to account for P/Read's observed  December morphology.
Given the implausibility of two separate impact-triggered emission events
in such short succession, continuous emission over several weeks between
August and October 2005 is the most likely explanation for the bi-directional
nature of P/Read's cometary activity as observed in December 2005.  As the
most plausible explanation for continuous emission is sublimation-driven
dust ejection, we therefore conclude that P/Read's activity is in fact
cometary in nature.

Finally, our modeling of P/Read's appearance on 2005 December 25 has the
added benefit of providing us with our best constraint on the start time of
the dust emission.  In November 2005, dust emitted from P/Read extends
exclusively to the southwest, leaving the observed tail length as our only
constraint.  Our best detection of the dust tail was obtained on 2005
November 26 when we observed the tail extending as far as 3 arcmin from the
nucleus before it reached the edge of our detector's field of view.  For the
fastest, smallest particles emitted by P/Read ($\beta=\beta_{max}\approx0.1$),
a tail of that length can be produced in approximately 50 days, constraining
the emission start time to 2005 October 7 (or earlier), or slightly more than
two weeks prior to its discovery on 2005 October 24.  Our December 2005 data
gives us a much stronger constraint, however, given that our modeling
demonstrates that dust particles of any size extending to the southwest
must have been emitted at least 120 days prior to observations, or around
2005 August 26 ({\it cf}. Figure~\ref{impulse_model}).  Unfortunately, it is
not possible to identify an upper limit to the total elapsed
emission time for P/Read, as small particles with very early emission times
simply become too diffuse to detect by the time of our observations,
while large particles with early emission times cannot be distinguished
from small particles with later emission times.

\section{DISCUSSION\label{discussion}}

Even though P/Read does not have orbital elements that place it
directly within the Themis asteroid family as do 133P's and 176P's orbital
elements (P/Read's eccentricity is slightly
higher than the general eccentricity range of the Themis family;
{\it cf}. Table~\ref{orbelem}), P/Read's orbital similarity to the other
known MBCs is striking.  Despite the fact that P/Read was
discovered serendipitously,
of roughly $4.2\times10^5$ asteroids tabulated as
of 18 August 2008 by the IAU Minor Planet Center, only about 70
(0.02\%) have values of $a$, $e$, and $i$ as close to 133P as does P/Read.
The orbital similarity among all three MBCs suggests that they
may be related in origin, either as fragments of the
recent break-up of a single icy parent body (which may or may not have
been a member of the Themis family),
or as fragments of the initial break-up of the Themis parent body
\citep[$\sim1\times10^8-2\times10^9$~yrs ago;][]{mar95}.
Intriguingly, 133P has in fact been associated with a recent break-up in
the asteroid belt that formed the Beagle family which is
thought to be $\lesssim$10~Myr old \citep{nes08}.  This newly-identified
family does not include P/Read or 176P ({\it cf}. Table~\ref{orbelem}),
however, and in any case, 
as discussed above, subsurface ice on the MBCs protected by no
more than a few meters of dust should be stable over the age of the solar
system \citep{sch08}.  As such, the existence of MBC ice alone does not
require the MBCs to be recently-produced fragments of larger bodies, or even
fragments of larger bodies at all.
The MBCs could simply be icy but otherwise ordinary outer
main-belt asteroids that have been individually collisionally ``activated''
\citep[{\it cf.} 133P;][]{hsi04}.

Dynamical considerations aside, P/Read's unusually strong activity
unequivocally makes it unique among the MBCs. As we argued for 133P, the
months-long duration of P/Read's dust emission is most consistent with the
sublimation of volatile ices, suggesting that it is a bona fide comet
\citep{hsi04}.  Unlike 133P and 176P, however, which never exhibit any
significant coma \citep{hsi06b}, P/Read displays a substantial coma, which
is optically dominated by significantly larger particles
($a\gtrsim100~\mu$m) than those ejected from 133P ($a\sim1-20~\mu$m).
The estimated terminal ejection velocities ($v_{ej}\sim0.2-3$~m~s$^{-1}$)
of the dust particles in P/Read's coma are comparable to those determined
for 133P ($v_{ej}\sim1-2$~m~s$^{-1}$), although the much smaller size of
P/Read ($r_e\sim0.3$~km) means that its escape velocity
($v_{esc}=(8G\rho\pi r^2/3)^{1/2}\sim0.2$~m~s$^{-1}$, assuming a bulk
density of $\rho=1000$~kg~m$^{-3}$) is about an order of magnitude
smaller than 133P's escape velocity ($v_{esc}\sim2$~m~s$^{-1}$).
This lower escape velocity would permit
more particles to escape, perhaps partly explaining P/Read's
much stronger dust emission and large coma particles.

The unusual strength of P/Read's activity could potentially also be
explained if P/Read experiences significantly different temperature conditions
from those experienced by 133P and 176P.  A much higher perihelion
temperature could cause more vigorous sublimation, while a much lower
equilibrium temperature over its entire orbit could allow P/Read to
preserve significantly more ice than the other MBCs.
While P/Read's slightly higher eccentricity does bring it closer to
the Sun at perihelion ($q_{\rm P/Read}=2.36$~AU) than the other MBCs
($q_{\rm 133P}=2.64$~AU; $q_{\rm 176P}=2.58$~AU), its surface temperature
(assuming a thermally-equilibrated graybody)
at perihelion is essentially equal to those of the other MBCs
($T_{\rm P/Read}(q)\approx184$~K, $T_{\rm 133P}(q)\approx174$~K,
and $T_{\rm 176P}(q)\approx176$~K),
meaning that peak surface temperature is unlikely to be a significant factor
in explaining P/Read's unusually vigorous activity.
Likewise, P/Read's only slightly lower equilibrium aphelion surface temperature
($T_{\rm P/Read}(Q)\approx142$~K, $T_{\rm 133P}(Q)\approx148$~K,
$T_{\rm 176P}(Q)\approx145$~K) and virtually identical average equilibrium
temperature over its entire orbit ($T_{\rm P/Read}(a)\approx159$~K,
$T_{\rm 133P}(a)\approx159$~K, $T_{\rm 176P}(a)\approx158$~K)
as compared to the other MBCs means that it is unlikely to be significantly
more icy than 133P or 176P, particularly considering its much smaller size.

Perhaps the most likely additional explanation for the strength of P/Read's
dust emission could be that the comet was activated much more recently than
the other MBCs.  In our current model, currently-active MBCs are thought to
have been recently activated by impacts that excavated sub-surface
reservoirs of volatile ice, exposing them to direct solar heating
\citep{hsi06b}. These impacts only trigger the activity, however, not sustain
it.  Instead, the dust emission of the MBCs is thought to be driven
by the sublimation of these newly-exposed patches of volatile material
and primarily modulated by seasonal fluctuations in the solar
illumination of the active sites (instead of increases in overall temperature
during perihelion passages as with other comets)
\citep{hsi04,hsi06a}.  In this way, a single impact on a body can
be responsible for multiple episodes of dust emission lasting several
months each time, long after the ejecta from that impact has dissipated.
This hypothesis is supported by the behavior of 133P, which has now been
observed to show activity on three separate occasions
in 1996, 2002, and 2007 \citep{els96,hsi04,jew07}, occupying
roughly the same portion of its orbit (the quadrant following perihelion)
each time.  We note that under this hypothesis, the apparent trend of all
three MBCs displaying activity near perihelion is most likely attributable
to observational bias, faint activity being much
more difficult to detect at larger heliocentric and geocentric distances.

As discussed in Sections~\ref{results} and \ref{dustmodeling}, however,
P/Read's estimated mass loss rate is approximately an order of magnitude
larger than that estimated for 133P, with P/Read's dust coma also constituting
a far larger percentage of the comet's total brightness than 133P's dust trail.
The relative weakness of 133P's activity may reflect the near-depletion
of its reserve of volatile material after multiple outbursts, the
small size of that reserve in the first place, or a combination of both.
Conversely, the more intense activity of P/Read in 2005 could indicate not
only the exposure of a larger reservoir of volatile material, but also
the more recent excavation of that reservoir, suggesting that we
might even be witnessing the immediate aftermath of an
activating impact on P/Read.
A particularly disruptive impact and exceptionally prodigious mass loss
that could be associated with the new exposure of a large volatile
reservoir might also explain the surprisingly small nucleus observed in 2007,
over a year after P/Read was observed to be active.
We caution that no evidence of fragmentation was observed
in either 2005 or 2007, but nonetheless,
this possibility is an important one to consider, given that unlike 133P's
activity, P/Read's activity has not yet been shown to be recurrent.
We emphasize that numerical models show that an impact is unlikely to
be the sole cause of P/Read's
dust emission, suggesting that it is driven by the sublimation of volatile
ice.  The combination of a triggering impact and the subsequent
vigorous sublimation that might be expected from a large, newly-exposed
reservoir of volatile material, however, could produce a particularly
strong initial burst of activity, but perhaps dramatically weaker future
outbursts, if any occur at all.

The key to an improved understanding of P/Read is to determine
whether it displays activity
similar to its 2005 outburst following its next perihelion
passage, or if any recurring activity is
significantly diminished in intensity (perhaps to a 133P- or 176P-like level).
Future observations to search for renewed dust emission near P/Read's next
perihelion passage on 2011 March 10, and also to assess its levels of
activity or inactivity in the months and years
prior to this date, will be needed to clarify these issues.

\section{SUMMARY}

Observations of the main-belt comet P/2005 U1 (Read) were obtained on multiple
occasions in 2005 and 2007.  Key results are as follows:
\begin{enumerate}
\item{Using data from 2007 (when P/Read appears to be largely inactive),
  we find an approximate absolute magnitude of $H_R\sim20.1\pm0.4$,
  corresponding to an effective radius of $r_e\sim0.3$~km (for an assumed
  albedo of $p_R=0.04$).
  }
\item{Monte Carlo numerical simulations of P/Read's dust emission indicate
  that the coma and tail are optically dominated by dust particles
  greater than 10~$\mu$m, and possibly greater than 100~$\mu$m in radius,
  with terminal ejection velocities of $v_{ej}\sim0.2-3$~m~s$^{-1}$.
  While these terminal ejection velocities are comparable to those found
  for 133P, the optically-dominant particle sizes are significantly larger
  for P/Read than for 133P.  P/Read's mass loss rate during its outburst
  is estimated to be $dm/dt\sim0.2$~kg~s$^{-1}$, roughly an order of
  magnitude larger than that estimated for 133P.
  }
\item{Optical colors of P/Read's dust coma
  are approximately solar and consistent with colors measured
  previously for 133P, other active and inactive comets, and C-type asteroids.
  }
\item{Impulsive ejection of dust ({\it e.g.}, by impact) is unable to
  account for the observed longevity of the coma and tail in 2005.  A
  sustained, continuous mechanism for dust ejection, likely the sublimation
  of exposed ice, is required.  Emission is determined to have begun at
  least two months prior to the discovery of activity, though we note that
  this is only a lower limit and that earlier start times are also possible.
  }
\item{We note that the activity of P/Read is much stronger than that of the
  other two MBCs.  We suggest that this may indicate that the impact assumed
  to have triggered P/Read's activity occurred
  very recently, and encourage observations near its next
  perihelion passage (2011 Mar 10) to search for significantly weaker
  emission that may confirm this hypothesis.
  }
\end{enumerate}

\begin{acknowledgements}
We thank Richard Wainscoat and Rita
Mann for donated telescope time, Fabrizio Bernardi and David Tholen for
accurate astrometry of Keck data, and Ian Renaud-Kim, Dave
Brennen, and John Dvorak at the UH 2.2~m,
Kathy Roth, Chadwick Trujillo, and Tony Matulonis at Gemini, and Greg
Wirth, Cynthia Wilburn, and Gary Punawai at Keck for their
assistance with our observations. We appreciate support of
this work from NASA through a planetary astronomy grant to DJ.
\end{acknowledgements}

\begin{deluxetable}{llcrcccrrrrrrr}
\scriptsize
\tablewidth{0pt}
\tablecaption{Observations of P/2005 U1 (Read)\label{obs_read}}
\tablecolumns{14}
\tablehead{
\colhead{UT Date} & \colhead{Tel.\tablenotemark{a}}
   & \colhead{$N$\tablenotemark{b}}
   & \colhead{$t$\tablenotemark{c}}
   & \colhead{Filters} & \colhead{$m_R$\tablenotemark{d}}
   & \colhead{$\theta_s$\tablenotemark{e}}
   & \colhead{$\nu$\tablenotemark{f}}
   & \colhead{$R$\tablenotemark{g}}
   & \colhead{$\Delta$\tablenotemark{h}}
   & \colhead{$\alpha$\tablenotemark{i}}
   & \colhead{$\alpha_{pl}$\tablenotemark{j}}
   & \colhead{$pa_{-\odot}$\tablenotemark{k}}
   & \colhead{$pa_{-v}$\tablenotemark{l}}
}
\startdata
2005 Jul 28 & \multicolumn{6}{l}{\it Perihelion}
  ...............................................................
        &   0.0 & 2.37 & 2.28 & 25.2 & 0.5 & 252.3 & 250.9 \\
2005 Nov 10 & UH2.2    & 32 &  9600 & $BVRI$     & 19.28$\pm$0.05
  & 1.0 &  31.4 & 2.44 & 1.45 &  0.6 &   0.1 & 258.6 & 253.0 \\
2005 Nov 19 & UH2.2    & 62 & 18600 & $VRI$      & 19.34$\pm$0.05
  & 0.9 &  34.0 & 2.45 & 1.47 &  3.8 & --0.1 &  73.6 & 252.4 \\
2005 Nov 20 & UH2.2    & 22 &  6600 & $VRI$      & 19.46$\pm$0.05
  & 0.9 &  34.2 & 2.45 & 1.47 &  4.3 & --0.1 &  73.6 & 252.4 \\
2005 Nov 21 & UH2.2    & 42 & 12600 & $BVRI$     & 19.37$\pm$0.05
  & 0.7 &  34.5 & 2.45 & 1.48 &  4.8 & --0.1 &  73.6 & 252.3 \\
2005 Nov 22 & UH2.2    & 16 &  4800 & $R$        & 19.28$\pm$0.05
  & 0.8 &  34.8 & 2.45 & 1.48 &  5.3 & --0.1 &  73.6 & 252.3 \\
2005 Nov 26 & Gem.   & 10 &  1110 & $g'r'i'z'$ & 19.72$\pm$0.10
  & 0.5 &  35.9 & 2.46 & 1.50 &  7.1 & --0.2 &  73.5 & 252.0 \\
2005 Dec 24 & UH2.2    &  6 &  1800 & $R$ & 20.12$\pm$0.03
  & 1.1 &  43.7 & 2.50 & 1.74 & 17.1 & --0.5 &  72.9 & 251.3 \\
2005 Dec 25 & UH2.2    &  6 &  1800 & $R$ & 20.16$\pm$0.03
  & 1.1 &  43.9 & 2.51 & 1.75 & 17.4 & --0.5 &  72.9 & 251.3 \\
2007 Jan 27 & Keck     &  4 &   720 & $R$ & 24.9$\pm$0.4
  & 0.6 & 123.0 & 3.43 & 2.49 &  5.2 & --0.5 & 284.6 & 290.1 \\
2008 May 19 & \multicolumn{6}{l}{\it Aphelion}
  .................................................................
        & 180.0 & 3.96 & 3.28 & 11.8 & 0.0 & 114.0 & 294.2 \\
2011 Mar 10 & \multicolumn{6}{l}{\it Perihelion}
  ...............................................................
        & 0.0 & 2.36 & 3.28 &  7.9 & --0.3 & 68.3 & 245.9 \\
\enddata
\tablenotetext{a} {Telescope used (UH2.2: University of Hawaii 2.2~m
		   telescope; Gem.: 8~m Gemini North telescope; Keck:
		   10~m Keck I Observatory)}
\tablenotetext{b} {Number of images}
\tablenotetext{c} {Total effective exposure time}
\tablenotetext{d} {Mean $R$-band magnitude of nucleus and coma inside
                    $4\farcs0$ (radius) aperture}
\tablenotetext{e} {FWHM seeing in arcsec}
\tablenotetext{f} {True anomaly in degrees}
\tablenotetext{g} {Median heliocentric distance in AU}
\tablenotetext{h} {Median geocentric distance in AU}
\tablenotetext{i} {Solar phase angle in degrees}
\tablenotetext{j} {Orbit plane angle (between the observer and object orbit
                   plane as seen from the object) in degrees}
\tablenotetext{k} {Position angle of the antisolar vector, as projected in the
                   plane of the sky, in degrees east of north}
\tablenotetext{l} {Position angle of the negative velocity vector, as projected
                   in the plane of the sky, in degrees
                   east of north}
\end{deluxetable}

\begin{deluxetable}{lcccl}
\footnotesize
\tablewidth{0pt}
\tablecaption{Colors Compared\label{colors}}
\tablecolumns{5}
\tablehead{
\colhead{} & \colhead{$B-V$}
  & \colhead{$V-R$} & \colhead{$R-I$} & \colhead{References}}
\startdata
  P/2005 U1 (Read)  & $0.63\pm0.05$ & $0.37\pm0.04$ & $0.39\pm0.04$ & This work \\
  Solar             & $0.67$        & $0.36$        & $0.35$        & \citet{har82,har90} \\
  133P/Elst-Pizarro (MBC) & $0.69\pm0.02$ & $0.41\pm0.03$ & $0.27\pm0.03$ & \citet{hsi04} \\
  46P/Wirtanen (active) & --- & $0.18\pm0.17$ & $0.39\pm0.16$ & \citet{epi99} \\
  47P/Ashbrook-Jackson (active) & --- & $0.36\pm0.23$ & $0.19\pm0.31$ & \citet{low03} \\
  103P/Hartley 2 (active)       & --- & $0.32\pm0.12$ & ---           & \citet{low03} \\
  Inactive comet nuclei\tablenotemark{a} & --- & $0.45\pm0.02$ & --- & \citet{jew02} \\
\enddata
\tablenotetext{a} {Average of 12 inactive comet nuclei ranging from $V-R=0.31\pm0.02$
  to $V-R=0.58\pm0.02$}
\end{deluxetable}

\begin{deluxetable}{lll}
\footnotesize
\tablewidth{0pt}
\tablecaption{Parameters used to model P/2005 U1 (Read) dust emission
               \label{fpmodel_params}}
\tablecolumns{3}
\tablehead{
\colhead{Parameter}
  & \colhead{Value(s)\tablenotemark{a}} & \colhead{Best-Fit\tablenotemark{b}}}
\startdata
$u_1$\tablenotemark{c} & 0.50 & 0.50 \\
$u_2$\tablenotemark{d} & 0.50 & 0.50 \\
$q$\tablenotemark{e}   & 3.5  & 3.5  \\
$k$\tablenotemark{f}   & 3.0  & 3.0  \\
$\beta_{max}$\tablenotemark{g} & $5\times10^{-1}$, $1\times10^{-1}$,
                                   $5\times10^{-2}$
                               & $1\times10^{-1}-5\times10^{-2}$ \\
$\beta_{min}$\tablenotemark{h} & $1\times10^{-5}$, $1\times10^{-4}$,
                 $1\times10^{-3}$, $1\times10^{-2}$ & 
                 $1\times10^{-4}$ \\
$v_0$\tablenotemark{i} & 10, 25, 50, 100, 150, 200 & 25 \\
$w$\tablenotemark{j} & 45$^{\circ}$, 90$^{\circ}$, 180$^{\circ}$
                 & 45$^{\circ}$ \\
\enddata
\tablenotetext{a} {Parameter values tested}
\tablenotetext{b} {Parameter values from among those tested that produced
                     simulated model images that provide the best match to
		     observed data}
\tablenotetext{c} {Power index of dependence of ejection velocity, $v_0$,
                     on $\beta$}
\tablenotetext{d} {Power index of dependence of ejection velocity, $v_0$,
                     on heliocentric distance, $r_h$}
\tablenotetext{e} {Power index of exponential size distribution of dust grains}
\tablenotetext{f} {Power index of dependence of dust production rate on
                     heliocentric distance, $r_h$}
\tablenotetext{g} {Maximum value in $\beta$ range tested}
\tablenotetext{h} {Minimum value in $\beta$ range tested}
\tablenotetext{i} {Ejection velocity in m~s$^{-1}$ of particles with
                     $\beta=1$ at a heliocentric distance of $r_h=1$~AU}
\tablenotetext{j} {Half-opening angle in degrees with respect to the
                     Sun-comet axis of assumed cone-shaped jet of dust
		     emission}
\end{deluxetable}

\begin{deluxetable}{lcrccccc}
\scriptsize
\tablewidth{0pt}
\tablecaption{$Af\rho$ measurements of observations and models of P/2005 U1 (Read)\label{afrho}}
\tablecolumns{8}
\tablehead{
   & \colhead{Days Since} &  &  &  
   & $Af\rho(\alpha=0)$ & $Af\rho(\alpha=0)$ & $Af\rho(\alpha=0)$ \\
\colhead{UT Date}
   & \colhead{Perihelion\tablenotemark{a}}
   & \colhead{$\alpha$\tablenotemark{b}}
   & \colhead{$m_{avg}(\alpha)$\tablenotemark{c}}
   & \colhead{$m_{avg}(\alpha=0)$\tablenotemark{d}}
   & \colhead{(obs.)\tablenotemark{e}}
   & \colhead{(contin.)\tablenotemark{f}}
   & \colhead{(termin.)\tablenotemark{g}}
}
\startdata
2005 Nov 10 & 105 &  0.6 & 19.28 & 19.26 & $7.86\pm0.39$ &  7.86 & 7.86 \\
2005 Nov 19 & 114 &  3.8 & 19.34 & 19.21 & $8.43\pm0.42$ &  9.85 & 7.41 \\
2005 Nov 20 & 115 &  4.3 & 19.46 & 19.31 & $7.67\pm0.38$ &  ---  & ---  \\
2005 Nov 21 & 116 &  4.8 & 19.37 & 19.20 & $8.52\pm0.43$ &  ---  & ---  \\
2005 Nov 22 & 117 &  5.3 & 19.28 & 19.09 & $9.41\pm0.47$ &  ---  & ---  \\
2005 Nov 26 & 121 &  7.1 & 19.72 & 19.47 & $6.79\pm0.68$ & 10.07 & 6.29 \\
2005 Dec 24 & 149 & 17.1 & 20.12 & 19.52 & $7.77\pm0.22$ &  8.16 & 3.46 \\
2005 Dec 25 & 150 & 17.4 & 20.16 & 19.55 & $7.67\pm0.22$ &  ---  & ---  \\
\enddata
\tablenotetext{a} {Days elapsed since the most recent perihelion passage on 2005 Jul 28}
\tablenotetext{b} {Solar phase angle in degrees}
\tablenotetext{c} {Average magnitude measured inside an aperture with a
                    $4\farcs0$ radius}
\tablenotetext{d} {Phase-angle-corrected average measured magnitude, using
                    $m_{avg}(\alpha=0) = m_{avg}(\alpha) - 0.035\alpha$ }
\tablenotetext{e} {$Af\rho$ in cm, calculated using $m_{avg}(\alpha=0)$}
\tablenotetext{f} {$Af\rho$ in cm, calculated for modeled comet image using
                     model for which
                     emission is continuous throughout the observation period,
		     normalized to observed $Af\rho$ on 2005 Nov 10}
\tablenotetext{g} {$Af\rho$ in cm, calculated for modeled comet image using
                     model for which emission is terminated as of 2005 Nov 10,
		     normalized to observed $Af\rho$ on 2005 Nov 10}
\end{deluxetable}

\begin{deluxetable}{lccccc}
\footnotesize
\tablewidth{0pt}
\tablecaption{Orbital Elements Compared
               \label{orbelem}\tablenotemark{a}}
\tablecolumns{6}
\tablehead{
\colhead{} & \colhead{$a$\tablenotemark{b}}
  & \colhead{$e$\tablenotemark{c}} & \colhead{$i$\tablenotemark{d}}
  & \colhead{$T_J$\tablenotemark{e}}
  & \colhead{$P_{orb}$\tablenotemark{f}}}
\startdata
133P/Elst-Pizarro       & 3.164 & 0.153 & 1.38 & 3.185 & 5.62 \\
P/2005 U1 (Read)        & 3.165 & 0.253 & 1.27 & 3.153 & 5.63 \\
176P/LINEAR             & 3.218 & 0.144 & 1.40 & 3.173 & 5.71 \\
Themis family\tablenotemark{g} & 3.05--3.22 & 0.12--0.19 & 0.69--2.23 & \\
Beagle family\tablenotemark{h} & 3.15--3.17 & 0.15--0.16 & 1.30--1.41 & \\
\enddata
\tablenotetext{a} {Elements shown for 133P and 176P are proper elements
                   from the AstDys website; elements for P/Read are
		   osculating elements from JPL's online database}
\tablenotetext{b} {Semimajor axis in AU}
\tablenotetext{c} {Eccentricity}
\tablenotetext{d} {Inclination in degrees}
\tablenotetext{e} {Tisserand invariant}
\tablenotetext{f} {Orbital period in years}
\tablenotetext{g} {Approximate orbital element bounds of the Themis
                    family \citep{zap90}}
\tablenotetext{h} {Approximate orbital element bounds of the Beagle
                    family \citep{nes08}}
\end{deluxetable}

\clearpage

\end{document}